\documentclass[11pt,a4paper]{article}

\usepackage[T1]{fontenc}
\usepackage[utf8]{inputenc}
\usepackage{lmodern}
\usepackage[margin=1in]{geometry}
\usepackage{amsmath,amssymb}
\usepackage{booktabs}
\usepackage{enumitem}
\usepackage{microtype}
\usepackage{tabularx}
\usepackage{xcolor}
\usepackage{hyperref}
\usepackage[round,authoryear,sort&compress]{natbib}
\usepackage{tikz}
\usetikzlibrary{arrows.meta,positioning}

\newcommand{\system}{\textsc{Incipit}}
\newcommand{\framework}{\textsc{Literary Axioms}}
\newcolumntype{Y}{>{\raggedright\arraybackslash}X}

\hypersetup{
  colorlinks=true,
  linkcolor=blue!55!black,
  citecolor=blue!55!black,
  urlcolor=blue!55!black,
  pdftitle={Incipit: Axiom-Grounded Scaffolding for Human-AI Literary Creation},
  pdfauthor={Qiang Liu and Chunyi Zhao},
  pdfsubject={Design and implementation of an axiom-grounded human-AI literary creation system},
  pdfkeywords={AI-assisted writing, creativity support tools, human-AI co-creation, narrative planning, knowledge grounding}
}

\title{\textbf{Incipit: Axiom-Grounded Scaffolding\\for Human--AI Literary Creation}}

\author{
  Qiang Liu\thanks{Corresponding author: \texttt{will.liuqiang@gmail.com}}\\
  \textit{Noevara Inc.}
  \and
  Chunyi Zhao\\
  \textit{Centre of Educational Design and Learning, University of Otago}\\
  \texttt{cccchunyi07@gmail.com}
}

\date{September 2026}

\begin{document}
\maketitle

\begin{abstract}
Large language models can produce fluent prose from short prompts, but a direct
prompt-to-text interaction gives writers little access to the assumptions that
shape a long narrative. We present \system{}, an implemented research prototype
that inserts an explicit planning layer between a writer's intent and generated
prose. The layer is grounded in \emph{literary axioms}: curated, reusable
propositions about human experience and narrative craft. The prototype connects
a knowledge base of 1,455 axioms and 472 typed relationships to a five-round
direction dialogue, a retrieval-and-selection pipeline, and a three-level
blueprint covering creative premises, story beats and character arcs, and
chapter outlines. Writers can inspect and edit the resulting structures before
using them as context for scene generation. Additional modules support
real-event abstraction and five-dimensional diagnostic feedback. We describe
the design rationale, data flow, implementation boundaries, and a worked design
example. Because no controlled user study or independently rated output study
has yet been completed, we do not claim that the system improves literary
quality. Instead, we outline a future preregistered comparison designed to distinguish
the contribution of axiom grounding from that of hierarchical planning. The
paper contributes a concrete architecture for making literary knowledge an
inspectable coordination object in human--AI writing.
\end{abstract}

\noindent\textbf{Keywords:} AI-assisted writing; creativity support tools;
human--AI co-creation; narrative planning; knowledge grounding; literary axioms

\section{Introduction}
\label{sec:introduction}

Generative language models have lowered the cost of producing plausible prose.
That capability does not by itself resolve a central problem in long-form
writing: decisions about conflict, character, theme, pacing, and point of view
must remain mutually intelligible across many scenes. In a direct
prompt-to-text interface, those decisions are typically compressed into a
prompt or left implicit in model parameters. The writer can revise the output,
but has limited means to inspect or negotiate the structural assumptions that
produced it.

Research systems have therefore explored mixed-initiative writing and explicit
planning. Creative Help supplies contextual continuations
\citep{roemmele2015creative}; Wordcraft studies co-writing interactions with
writers \citep{yuan2022wordcraft}; TaleBrush provides a graphical
control for a story's emotional trajectory \citep{chung2022talebrush}; and
Dramatron decomposes script generation into log lines, characters, locations,
beats, and dialogue \citep{mirowski2023cowriting}. These systems demonstrate the
value of intermediate representations. A remaining design question is what
kind of knowledge should populate such representations, and how writers can
inspect its influence on the developing work.

We investigate that question through \system{}, a full-stack research prototype
for axiom-grounded literary creation. Following the companion framework, an
``axiom'' is a provisional organizing premise, represented in the system as a
reusable proposition in a curated knowledge resource. It does
not denote a self-evident truth or carry a mathematical truth guarantee. An
axiom may express a thematic observation or a craft principle; it may also be
culturally situated, contestable, or productive precisely because it is in
tension with another proposition. The companion paper defines this framework
and documents its dataset \citep{liu2026axioms}. The present paper asks:

\begin{quote}
How can explicit propositions about literature serve as inspectable scaffolding
between a writer's intent and model-generated narrative material?
\end{quote}

The system operationalizes this question as a sequence: elicit intent, retrieve
candidate axioms, let the writer establish a working set, generate a
hierarchical blueprint, and use the blueprint during scene-level writing and
diagnosis. Its central design commitment is procedural rather than empirical:
\emph{structure precedes prose generation}, while the writer can intervene at
each structural level.

This paper makes four contributions:

\begin{enumerate}[leftmargin=*]
  \item an interaction model in which curated literary propositions act as a
  shared, inspectable vocabulary between writer and model;
  \item an implemented five-round direction dialogue and five-stage matching
  pipeline that convert an initial idea into a selected axiom set;
  \item an implemented three-level blueprint that carries selected propositions
  into editable creative premises and subsequent planning context, with partial
  tracing across levels; and
  \item a proposed evaluation protocol for distinguishing the contribution of
  axiom grounding from that of dialogue and hierarchical planning.
\end{enumerate}

Our contribution is a system design and implementation account. The worked
example in Section~\ref{sec:walkthrough} is illustrative and is not an outcome
evaluation.

\section{Related Work}
\label{sec:related}

\subsection{Creativity support and mixed initiative}

Creativity support tools should help users explore alternatives, relate ideas,
compose artifacts, and retain control over consequential decisions
\citep{shneiderman2007creativity}. Mixed-initiative systems distribute action
between human and computational participants rather than treating one as a
passive operator \citep{yannakakis2014mixed,deterding2017mixed}. The COFI
framework further characterizes co-creative interaction through participation,
communication, contribution, and turn-taking \citep{rezwana2023designing}.
These accounts motivate our treatment of the blueprint as a coordination
artifact: the model contributes candidates and structure, while the writer can
accept, edit, replace, or reject them.

Human--AI interaction guidelines also emphasize making system capabilities
clear, supporting correction, and preserving user control
\citep{amershi2019guidelines}. In writing, these requirements are difficult to
satisfy when the only visible artifacts are a prompt and a generated passage.
\system{} exposes intermediate decisions, although the current prototype does
not yet explain every model inference or calibrate confidence.

\subsection{AI-assisted writing and narrative planning}

Story-writing tools have adopted several levels of control. Creative Help
offers sentence-level suggestions \citep{roemmele2015creative}. Wordcraft
supports free-form co-writing operations and documents how writers
appropriate them \citep{yuan2022wordcraft}. TaleBrush lets users sketch a
fortune curve to steer generation \citep{chung2022talebrush}. Dramatron uses a
hierarchical generation pipeline for dramatic scripts
\citep{mirowski2023cowriting}. Story generation research has also examined
planning as a means to improve long-range coherence
\citep{fan2018hierarchical,yao2019planwrite}.

\system{} shares the premise that intermediate structure is useful, but gives
the intermediate layer a curated semantic vocabulary. The writer can identify
which propositions support the core conflict or narrative technique, rather
than working only with generic beats or latent model associations.

\subsection{Narrative formalisms and explicit knowledge}

Narrative theory has long supplied reusable analytical concepts, including
character functions \citep{propp1968morphology}, temporal order and focalization
\citep{genette1980discourse}, and story grammars
\citep{rumelhart1975notes}. Computational creativity similarly distinguishes
exploration within a conceptual space from transformation of that space
\citep{boden2004creative}. These traditions show both the utility and the risk
of formalization: a representation can make patterns operational, but can also
present contingent interpretations as universal rules.

The \framework{} representation therefore includes contextual types and typed
relations, and treats axioms as prompts for deliberation rather than constraints
that a literary work must obey. This stance differs from a theorem-proving
interpretation of ``axiom'' and from an ontology that claims exhaustive coverage
of literary meaning.

\section{Knowledge and Design Commitments}
\label{sec:commitments}

\subsection{The knowledge resource}

The checked-in knowledge resource contains 1,455 axiom records, 1,464
work--axiom mappings referring to 149 unique works, and 472 inter-axiom
relationships. Axioms are divided into content and form categories, assigned a
context type and abstraction level, and organized across 11 domains. Relations
are typed as \texttt{complements}, \texttt{tensions}, \texttt{contradicts},
\texttt{specializes}, or \texttt{evolves\_from}. The companion paper gives the
construction procedure, descriptive statistics, and evidence limitations
\citep{liu2026axioms}.

For this system, an axiom record serves three functions. Its statement provides
a compact concept for retrieval and discussion. Its metadata supports filtering
and diversity checks. Its work mappings and relations provide context for
selection. None of these fields establishes that the statement is universally
valid, and the system permits selection of axioms in tension when that tension
is creatively deliberate.

\subsection{Four design commitments}

\paragraph{Make grounding visible.}
The system displays selected axioms as named creative cores and carries their
identifiers into the blueprint. This exposes part of the context used for
generation and gives the writer a concrete object to revise.

\paragraph{Move from coarse to fine structure.}
The system orders decisions from direction to premise, beats, chapters, and
scenes. Later stages receive earlier artifacts as context. This hierarchy is
intended to reduce accidental drift; whether it improves perceived coherence is
an empirical question addressed by the proposed evaluation.

\paragraph{Preserve human intervention points.}
The writer can supplement dialogue answers, choose among suggested options,
patch each blueprint layer, request alternative cores, regenerate an individual
chapter outline, and revise scene text. These are concrete controls in the
prototype. General blueprint branching and automatic version creation for every
edit are outside the current implementation.

\paragraph{Present diagnostics as advice.}
Automated scores and suggestions depend on model judgments and simple
heuristics. They are displayed as revision support, not as measurements of
literary merit. The interface should make this status explicit, particularly
for culturally dependent concepts such as universality and originality.

\section{Workflow and System Design}
\label{sec:system}

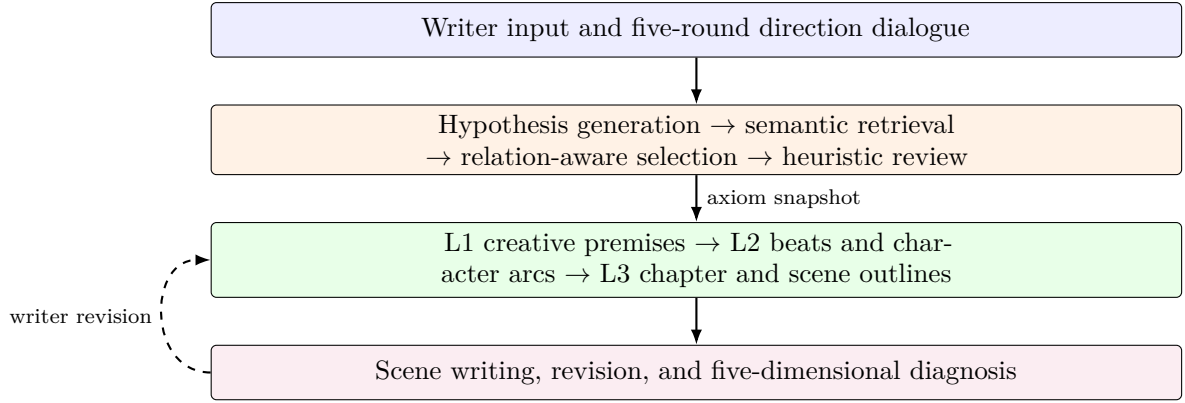
\begin{figure}[t]
\centering
\begin{tikzpicture}[
  node distance=0.62cm,
  box/.style={draw, rounded corners=2pt, align=center, minimum height=0.72cm,
              text width=0.79\linewidth, font=\small},
  arrow/.style={-{Latex[length=2mm]}, thick}
]
\node[box, fill=blue!7] (intent) {Writer input and five-round direction dialogue};
\node[box, fill=orange!10, below=of intent] (match) {Hypothesis generation $\rightarrow$
semantic retrieval $\rightarrow$ relation-aware selection $\rightarrow$ heuristic review};
\node[box, fill=green!9, below=of match] (bp) {L1 creative premises $\rightarrow$
L2 beats and character arcs $\rightarrow$ L3 chapter and scene outlines};
\node[box, fill=purple!7, below=of bp] (write) {Scene writing, revision, and five-dimensional diagnosis};
\draw[arrow] (intent) -- (match);
\draw[arrow] (match) -- node[right,font=\scriptsize] {axiom snapshot} (bp);
\draw[arrow] (bp) -- (write);
\draw[arrow, dashed] (write.west) to[out=180,in=180,looseness=1.45]
  node[left,font=\scriptsize,align=right] {writer\ revision} (bp.west);
\end{tikzpicture}
\caption{Core \system{} workflow. Dashed feedback denotes writer-led revision,
not automatic optimization.}
\label{fig:architecture}
\end{figure}

Figure~\ref{fig:architecture} shows the main data flow. The knowledge layer
supplies semantic candidates and typed relations. The workflow layer turns user
input into structured project state. The interaction layer exposes dialogue,
blueprint, and scene operations.

\subsection{Five-round direction dialogue}

The dialogue service stores five ordered rounds: story core, core conflict,
differentiation, expression style, and confirmation. The first round accepts
free text. Rounds two through four can present selectable options while also
accepting a custom response. The final round consolidates the creative seed and
starts the matching pipeline. Responses are stored with extracted structured
fields, allowing the accumulated context to inform the next round.

The dialogue endpoint uses server-sent events for progressive status and text
delivery where the underlying provider supports streaming. This is an
interaction mechanism; it does not imply that all generation endpoints are
token-streamed or interruptible.

\subsection{Five-stage axiom matching}

The matching service uses the following stages:

\begin{enumerate}[leftmargin=*]
  \item \textbf{Understand:} parse the input into a creative seed containing
  conflict, character cues, setting, tone, and thematic hints.
  \item \textbf{Hypothesize:} generate several thematic search hypotheses from
  the seed.
  \item \textbf{Retrieve:} embed the hypotheses, perform cosine-similarity
  search over axiom vectors, and optionally add candidates from domains implied
  by thematic hints.
  \item \textbf{Assemble:} give the model candidate records and relations among
  them; retain only identifiers present in the retrieved candidate set and
  assign primary, supporting, or form roles.
  \item \textbf{Review:} compute transparent heuristics for contradiction edges,
  context-type composition, overlap with mapped works, and coverage. When a
  heuristic is low, generate revision suggestions.
\end{enumerate}

The review values are navigation aids. For example, the current ``universality''
heuristic is the proportion of Type A axioms in a selection. A high value does
not demonstrate that a story is better or culturally universal. Likewise, the
``originality'' heuristic is inversely related to the largest work-level axiom
overlap; it cannot detect stylistic originality or plagiarism.

Once the writer confirms a direction, the selected records and creative seed
are stored in the project's axiom snapshot. This gives later prompts a project-
level trace, although regeneration can reload current database records and the
private artifact does not provide immutable versioning for every edit.

The matching stages separate interpretation, retrieval, selection, and review.
S1 and S2 use the model to interpret an idea; S3 searches the database; S4
checks selected axiom IDs against the candidate map; and S5 combines database
counts with model advice. An S2 hypothesis is not a knowledge-base record.
An S4 selection is a live database record, which can include an unreviewed
proposal from the recovery branch. Candidate schemas do not retain record
status at every stage, so provenance is not uniformly visible downstream.
Membership checks apply to selected axioms; proposed tension endpoints are
parsed as UUIDs without the same membership check.

For a concrete view of the retrieval contract, let $H=\{h_1,\ldots,h_m\}$ be
the hypotheses returned by S2 and let $R(h_i)$ be the at-most-five nearest
records returned for $h_i$ by the embedding query. The semantic candidate pool
is the union of these sets after de-duplication, augmented by at most three
records per keyword-implied domain when thematic hints match the current domain
vocabulary. Candidates are sorted by their best observed similarity and the
service retains at most 50 records. S4 sees the first 30 records and the
relations whose endpoints are both in the candidate pool. These are
implementation bounds, not claims about an optimal retrieval policy; they make
the context size and the possible selection space explicit.

The dialogue path attempts recovery when S3 fails, returns fewer than five
candidates, or its maximum score is below 0.3. The model can propose up to three
records. These are inserted with status \texttt{proposed}, assigned to an
existing domain, and submitted for embedding. Retrieval is retried when new
records were created or the first attempt failed. If retrieval still fails,
the service returns a seed-only evaluation. A weak but nonempty pool can
continue to S4 and need not lead to this fallback. Despite being called
``temporary'' in one helper name, proposals are persistent database records.
They initially lack curated work mappings and relationships. Retrieval excludes
deprecated records but can return proposed records; database membership is
therefore not evidence of prior human review.

\subsection{Three-level blueprint}

Table~\ref{tab:blueprint} summarizes the implemented schema. Each level is
stored as structured JSON and validated before persistence.

\begin{table}[t]
\centering
\small
\begin{tabularx}{\linewidth}{@{}p{0.10\linewidth}Y Y@{}}
\toprule
\textbf{Level} & \textbf{Primary fields} & \textbf{Connection to earlier state} \\
\midrule
L1 & creative cores, core conflict, characters and relations, setting, genre,
narrative choices, tone, themes & selected axiom IDs, statements, roles, and
mapped-work references \\
L2 & synopsis, structure type, ordered beats, emotional levels, turning points,
character arcs, estimated length & L1 premises and creative-core statements \\
L3 & chapters, scenes, point-of-view character, key events, beat IDs, emotional
arc, word budgets, writing notes & L2 beat IDs and character arcs; references
checked when the L2 beat set is nonempty \\
\bottomrule
\end{tabularx}
\caption{Implemented blueprint levels and their principal dependencies.}
\label{tab:blueprint}
\end{table}

The writer can patch all three levels. L1 generation limits the number of cores
according to the target-length profile and enriches mapped axioms with up to
three work references. L2 supports five structure labels: three-act, hero's
journey, in medias res, frame, and parallel. L3 distributes a target word budget
across chapters and scenes and checks supplied chapter beat references against
a nonempty L2 beat set. An endpoint can regenerate one chapter outline while preserving the
remainder. Confirming the blueprint materializes workshop characters,
structures, and scenes for subsequent writing.

The blueprint evaluator stores up to 20 evaluation snapshots. It combines
relation-based coherence, Type A proportion, work overlap, and completion of
the three layers. This evaluation history should not be confused with complete
blueprint version history: edits update the current blueprint record.

The hierarchy can be understood as a sequence of contracts rather than as a
claim that literary planning is intrinsically hierarchical. L1 establishes the
objects and commitments that later layers may refer to; L2 turns those
commitments into ordered change; L3 allocates that change to bounded writing
units. A layer is useful when it gives the next layer something that can be
checked. For example, a chapter's \texttt{beat\_ids} can be checked against L2,
whereas a free-form ``theme'' field cannot. Conversely, the contracts do not
guarantee that a valid reference is a good artistic choice. A writer can accept
a structurally valid but aesthetically unconvincing beat and then edit it.

\subsection{State transitions and intervention points}
\label{sec:state}

The persisted project state makes the system's intended ordering visible. A
creative session begins as \texttt{active}; each round stores the writer's
response and extracted fields; completion stores a seed and the pipeline result
in an axiom snapshot. A blueprint then moves through \texttt{draft},
\texttt{layer1}, \texttt{layer2}, \texttt{layer3}, and \texttt{confirmed}. The
status names are workflow markers, not quality labels. A failed JSON parse can
trigger one repair request, while a schema or cross-level validation failure
stops persistence and exposes the error to the caller.

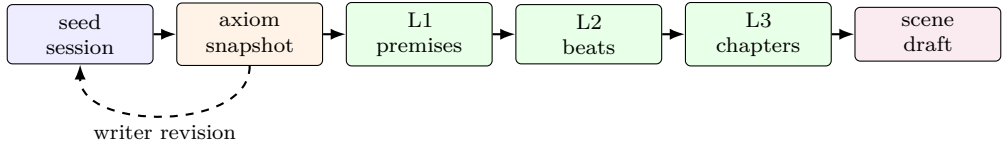
\begin{figure}[t]
\centering
\begin{tikzpicture}[
  node distance=0.25cm and 0.30cm,
  state/.style={draw, rounded corners=2pt, align=center, minimum height=0.58cm,
                text width=0.105\linewidth, font=\scriptsize},
  arrow/.style={-{Latex[length=2mm]}, thick}
]
\node[state, fill=blue!7] (seed) {seed\\session};
\node[state, fill=orange!10, right=of seed] (axiom) {axiom\\snapshot};
\node[state, fill=green!9, right=of axiom] (l1) {L1\\premises};
\node[state, fill=green!9, right=of l1] (l2) {L2\\beats};
\node[state, fill=green!9, right=of l2] (l3) {L3\\chapters};
\node[state, fill=purple!8, right=of l3] (scene) {scene\\draft};
\draw[arrow] (seed) -- (axiom);
\draw[arrow] (axiom) -- (l1);
\draw[arrow] (l1) -- (l2);
\draw[arrow] (l2) -- (l3);
\draw[arrow] (l3) -- (scene);
\draw[arrow, dashed] (axiom.south) to[out=-90,in=-90] node[below,font=\scriptsize] {writer revision} (seed.south);
\end{tikzpicture}
\caption{Persisted workflow states and principal intervention points. Dashed
edges indicate writer-led operations; they are not automatic search loops.}
\label{fig:state-machine}
\end{figure}

Figure~\ref{fig:state-machine} summarizes the intended transitions. Human
intervention is possible after each completed stage: the writer may revise the
seed, choose a different candidate set, patch a layer, or regenerate one
chapter. The current implementation does not create a general branching graph
of alternatives, so an edit updates the current record and the evaluator's
bounded history is not a complete provenance log.

Propagation is also partial. Editing L1 cores updates the blueprint but does
not synchronize the project's axiom snapshot. Confirmation copies characters,
chapter goals, and selected scene-planning fields into workshop records, but
does not carry over chapter \texttt{beat\_ids}, \texttt{core\_links}, or
\texttt{writing\_notes}. Core links within the blueprint are free-text references.
Scene generation and diagnosis continue to read the project snapshot;
consequently, edits to blueprint cores can diverge from the axiom context used
downstream. The implemented hierarchy provides inspectable planning artifacts,
not a fully synchronized provenance graph.

A confirmed blueprint is mutable, and generated prose is not evidence that the
selected axioms were realized. The snapshot provides partial tracing through
selected IDs and seed information. L1 generation reloads current statements and
work references, and evaluation consults current relationships. Exact prompt
replay would additionally require versioned database records, templates,
provider settings, model identifiers, and responses.

\subsection{Scene writing and diagnosis}

The writing workshop exposes scene discussion, generation, continuation, and
revision operations. Scene generation assembles the project axiom snapshot,
character information, scene goal, previous prose, and length guidance. The
writer agent also reads a chapter summary field when present in the selected structure;
blueprint confirmation does not populate that field. Server-sent events stream
discussion and generated scene content. In the reported workflow, prose
production follows blueprint confirmation, with the partial transfer described
above, while the text remains editable.

The diagnosis service evaluates a project excerpt along five prompt-defined
dimensions: self-consistency, universality, originality, consistent deduction,
and connotation. It returns per-dimension scores, findings, comparable-work
suggestions, and revision suggestions. Diagnosis reads at most the first 6,000
characters of project content, so it cannot establish whole-manuscript
consistency. Its classic-work reference context is empty. Moreover, the normal
pipeline's selected-record schema omits context type, leaving the diagnostic
type-distribution counts at zero for those snapshots. Scores and comparison
suggestions are model outputs and have not been
validated against expert judgments.

\subsection{Real-event abstraction}

An optional entry route transforms a description of a real event into proposed
story cores. Its prompt requests three operations: abstract identifying details
into thematic queries, migrate the setting across a fixed six-cell era--region
grid, and estimate residual identifiability. Retrieved story cores must refer to
existing axiom IDs before they are shown to the writer. The module can help a
writer consider distance from source material, but its model-produced
identifiability score is not a privacy guarantee. Publication still requires
human review for consent, defamation, confidentiality, and contextual harms.

\section{Implementation}
\label{sec:implementation}

\system{} is implemented as a web application with a Python/FastAPI backend and
a Next.js/React TypeScript frontend. PostgreSQL stores users, projects, dialogue
rounds, axiom records, typed relations, blueprints, scenes, and diagnoses.
Structured artifacts use PostgreSQL JSONB fields; embeddings use the pgvector
extension. Pydantic schemas validate generated blueprint objects, including
enumerated structure types, emotional-level ranges, and cross-level beat
references.

The provider registry supports OpenAI, Anthropic, and DeepSeek-compatible LLM
backends and OpenAI or Voyage embedding backends. Prompted JSON operations make
one repair request when the first response is not valid JSON. The code does not
implement automatic failover between providers. Consequently, the prototype's
design should be described as provider-configurable rather than provider-
redundant.

Vector dimensionality is a deployment constraint. The current ORM schema uses
1,024-dimensional vector columns, while the OpenAI provider exposes a
configurable dimension and the Voyage adapter reports 1,024 dimensions. A
deployment must choose a provider configuration that matches the database
schema or apply a coordinated schema migration and re-embed the corpus. This
paper reports neither a fixed production embedding configuration nor measured
retrieval latency.

Table~\ref{tab:traceability} links paper-level claims to implemented modules.
The mapping is included to make the system account auditable without treating
source-code existence as evidence of user benefit.

\begin{table}[t]
\centering
\small
\begin{tabularx}{\linewidth}{@{}Y Y Y@{}}
\toprule
\textbf{Capability} & \textbf{Implemented state} & \textbf{Primary module} \\
\midrule
Five-round dialogue & persisted rounds, option selection, custom text, SSE
delivery & \texttt{services/dialogue.py} \\
Axiom matching & understand, hypothesize, retrieve, assemble, review &
\texttt{services/pipeline.py} \\
Three-level blueprint & validated L1--L3 generation, editing, one-chapter
regeneration & \texttt{services/blueprint.py} \\
Five-dimensional diagnosis & excerpt-based prompted assessment and history &
\texttt{services/diagnosis.py} \\
Real-event abstraction & thematic extraction, migration grid, axiom retrieval &
\texttt{services/event.py} \\
\bottomrule
\end{tabularx}
\caption{Traceability from reported capabilities to the backend implementation.}
\label{tab:traceability}
\end{table}

\subsection{Prompt assembly and validation boundaries}
\label{sec:contracts}

The services treat model responses as structured input requiring validation.
Module-specific JSON helpers use a one-repair pattern after parsing fails.
The pipeline wraps a repeated failure in a stage-specific exception; the
blueprint helper can propagate a JSON decoding exception. These helpers are
not a single uniform error contract. Syntactically valid JSON can still encode
an unsuitable story.

Blueprint generation adds a second boundary through Pydantic models. L1 checks
the role of each creative core, normalizes character lists, and preserves the
axiom identifier supplied by the snapshot. L2 constrains structure labels,
beat-identifier syntax, positive order values, and emotional levels in $[0,1]$;
it does not enforce unique beat identifiers or agreement between list position
and order values. L3 constrains
chapter numbers, scene word counts, emotional levels, and the shape of chapter
arcs. L3 generation and patch endpoints check chapter \texttt{beat\_ids}
against the current L2 only when its beat set is nonempty; empty chapter
reference lists are permitted. Character-arc beat references are checked for
identifier syntax, not membership. Later L2 patches do not revalidate every existing L3
outline, so these checks do not guarantee persistent cross-level consistency.
They also cannot judge whether a scene is compelling.

Table~\ref{tab:contracts} records the main input, output, and failure boundary
for each generation stage. The table describes the implemented contract rather
than a recommended future API.

\begin{table}[t]
\centering
\small
\begin{tabularx}{\linewidth}{@{}p{0.15\linewidth}Y Y Y@{}}
\toprule
\textbf{Stage} & \textbf{Context assembled} & \textbf{Stage output} &
\textbf{Boundary} \\
\midrule
S1 & Raw direction text & Structured creative seed & JSON repair; seed fields
may still be inferred \\
S2 & Seed fields and dialogue context & 3--10 thematic hypotheses & Fewer than
three hypotheses raises a retryable stage error \\
S3 & Hypothesis embeddings and thematic hints & At most 50 de-duplicated
candidates & Weak retrieval enters the proposed-record recovery branch; a
retrieval failure after recovery permits seed-only evaluation \\
S4 & Candidate records and in-pool relations & Selected IDs, roles, tensions &
Selected axiom IDs outside the pool are discarded; fewer than two selections fails \\
S5 & Selected records, relation counts, and overlap & Scores and recommendations
& Heuristic scores are stored with model advice; no expert calibration \\
L1 & Seed, axiom snapshot, length profile & Premises, characters, setting &
Pydantic validation; axiom IDs are copied from the snapshot \\
L2 & L1 fields and length profile & Synopsis, beats, character arcs & Enumerated
structure type and bounded emotional levels \\
L3 & L1/L2 fields and target words & Chapters and scenes & Word-budget
normalization and conditional beat-membership check \\
\bottomrule
\end{tabularx}
\caption{Implemented generation contracts and their limits. A boundary marks
where the service stops or normalizes data; it is not a quality guarantee.}
\label{tab:contracts}
\end{table}

The table distinguishes stage outputs from durable records. The final pipeline
result stores seed, hypotheses, selected system, and evaluation; the retrieval
candidate pool is not retained there. A snapshot alone cannot replay selection.
System templates in \texttt{services/prompts.py} are English, while dynamic
context labels can be Chinese. Language policy controls generated content and
preserves machine-readable keys. Streaming support at dialogue and writing
endpoints does not imply that blueprint and diagnosis JSON calls are streamed,
interruptible, or reproducible.

\subsection{Planning and axiom grounding as separate variables}
\label{sec:planning-grounding}

The workflow contains two interventions that are easy to conflate. The first is
\emph{hierarchical planning}: asking a model to move from a seed to L1, L2,
and L3 before drafting prose. The second is \emph{axiom grounding}: supplying
curated propositions, their roles, relations, and mapped-work context while
making those references editable. A plan can be generated without the curated
knowledge layer, and an axiom can be shown without requiring the writer to
accept a particular beat sequence. The system's design therefore supports a
three-condition comparison rather than a single before/after claim.

The Plan and Axiom conditions would share dialogue, length profiles, schemas,
and common prompt components, with knowledge fields withheld in Plan. Their
contrast would estimate the incremental contribution of the visible knowledge
layer within this workflow. An additional exploratory comparison could remove
the blueprint while retaining axiom candidates. That axiom-only condition is
outside the three-arm protocol below and would require a separate specification.
The current prototype has not run these ablations.

The distinction matters for the five dimensions in the originating theory:
coherence, contextual reach, distinctiveness, realization fidelity, and
interpretive richness. They are treated here as contested hypotheses about what
readers may value, not as objective scores. Incipit exposes proxies for some of
them: the matching and blueprint evaluators use relation and contradiction
counts as a partial check on internal compatibility, Type A proportions as a
context heuristic, and work-level overlap as a narrow lineage signal; the
separate diagnosis prompts ask about deduction and connotation. None of these
proxies establishes the corresponding literary property. The proposed study
must therefore compare human ratings and writer experience against the proxies
rather than validating the proxies by their own outputs.

The software vocabulary should not be read as a one-to-one implementation of
the companion theory's five dimensions. In diagnosis, ``self-consistency''
addresses aspects of coherence, ``consistent deduction'' addresses aspects of
realization fidelity, and ``connotation'' prompts consideration of interpretive
richness. Diagnostic ``originality'' asks the model about thematic and formal
novelty, whereas matching and blueprint ``originality'' uses work overlap alone.
Diagnostic ``universality'' asks about abstraction levels and type diversity;
matching and blueprint ``universality'' instead uses the Type A proportion.
Neither assesses contextual reach across reader populations. These are partial,
unvalidated proxies: contradiction penalties do not distinguish accidental from
deliberately organized conflict, and thematic alignment does not establish
realization fidelity. The matching and blueprint evaluators additionally report
``completeness,'' a coverage indicator outside the five theoretical dimensions.
Software labels are retained to match the implementation; theoretical dimensions
and software proxies must be evaluated separately.

\section{Worked Design Example}
\label{sec:walkthrough}

This section demonstrates representational flow, not measured effectiveness.
The initial premise and intermediate artifacts are authored examples rather
than records from a user study or a frozen model run.

Suppose a writer begins with: ``A woman returns to her childhood town after two
decades and discovers that her memories no longer fit the place.'' During the
direction dialogue, the writer might choose loss of an idealized past as the
core conflict, distinguish the story by making the town's apparent continuity
rather than visible change unsettling, and choose a close third-person voice
with a restrained tone.

For illustration, suppose retrieval surfaces the record labelled
\texttt{ax\_001} in the checked-in resource. Using the same author translation
as the companion paper, its statement is: ``Desire ultimately pursues the
temporal illusion carried by its apparent object.'' The writer may accept it
as a primary core, reject it, or request alternatives. The dataset label is
used here for readability; imported records receive runtime UUIDs. If selected,
the record's UUID and statement enter the project snapshot.

Table~\ref{tab:example} shows one possible propagation through the blueprint.
The point is traceability: each later artifact can be inspected against the
selected proposition.

\begin{table}[t]
\centering
\small
\begin{tabularx}{\linewidth}{@{}p{0.13\linewidth}Y@{}}
\toprule
\textbf{Artifact} & \textbf{Illustrative content} \\
\midrule
Axiom & \texttt{ax\_001}: desire may be directed toward an illusion of recovered
time. \\
L1 & The protagonist seeks restoration of a remembered relationship; her
conflict is between preserving the memory and encountering the present person. \\
L2 & Early beats reward nostalgic interpretation; a midpoint contradiction
reveals that her memory omitted another character's sacrifice; the final turn
requires relinquishing restoration without denying the past. \\
L3 & Chapter outlines link the midpoint chapter to the relevant beat, identify
the point-of-view character, specify the emotional transition, and assign a
word budget. \\
Scene & A writing prompt receives a scene goal and the project axiom snapshot, while
the writer remains free to contradict the proposed realization. \\
\bottomrule
\end{tabularx}
\caption{Illustrative propagation of one dataset axiom through the blueprint.
Only the axiom record is drawn from the checked-in resource; the story artifacts
are design examples.}
\label{tab:example}
\end{table}

The example demonstrates how a compact proposition can coordinate decisions at
different scales. It does not show that the resulting story is more original,
coherent, or valuable than one written with another tool. Those claims require
the evaluation described next.

\subsection{From axiom to scene: a constructed trace}
\label{sec:constructed-trace}

To make the propagation more concrete, consider a deliberately short trace
as a variant of the same premise. The following material was written for this paper; it is
not a transcript of a production run and does not imply that a configured
provider would return these exact words. The trace is useful because it shows
where a writer can disagree with the scaffold.

\begin{description}[leftmargin=2.2cm,style=nextline]
  \item[Working proposition] \texttt{ax\_001} is treated as a hypothesis about
  desire and remembered time. The writer adds a note: ``Use the proposition to
  test the protagonist's interpretation, not to announce a lesson.''
  \item[L1 commitment] Mara, a municipal archivist, returns to the town where
  her sister disappeared. She wants the old house to confirm that the sisters'
  last summer was shared in the way she remembers. Her contradiction is that
  she distrusts public records while needing a record to authorize her memory.
  \item[L2 change] Beat 1 rewards the memory: a neighbor repeats a phrase Mara
  associates with her sister. Beat 3 introduces a ledger in which the phrase is
  attributed to Mara herself. Beat 5 forces a choice between preserving the
  consoling version and acknowledging the sister's unrecorded account.
  \item[L3 allocation] Chapter 2 contains the ledger discovery and assigns two
  scenes: a quiet archive search followed by a conversation that ends before
  either character states what the ledger means. Its chapter goal is to make
  the protagonist's certainty costly; its core link points back to the axiom
  title rather than copying the axiom statement into dialogue.
  \item[Scene draft] ``The ledger smelled of dust and river water. Mara found
  the sentence in her own hand, the ink tilted toward the margin. She read it
  twice before the room acquired a second door. On the other side, she imagined
  her sister waiting with the patience of someone who had never agreed to be
  remembered.''
\end{description}

The scene is intended to realize the proposition through an image and a changed relation
between record and memory; it does not prove that the proposition is true. A
writer might revise it by rejecting the archive entirely, changing the point of
view, or reconsidering how a tension in the axiom snapshot is developed. The current
diagnosis can flag a possible inconsistency or ask whether the passage supports
multiple interpretations, but it cannot determine whether the image is
artistically successful. Nor can the 6,000-character excerpt establish that the
ledger remains consistent with a later chapter. This is why the trace is a
design demonstration and why the evaluation protocol separates structural
checks from independent literary judgment.

\section{Evaluation Protocol}
\label{sec:evaluation}

No controlled user study, validated survey dataset, independently rated corpus,
or preregistered comparison is available for the present paper. We therefore
outline a protocol for future work and make no inferential claims from informal
use. Primary outcomes, contrast, sample size, and analysis details must be
fixed before preregistration and recruitment.

\subsection{Research questions and conditions}

The central comparison should isolate axiom grounding from hierarchical
planning:

\begin{description}[leftmargin=2.4cm,style=nextline]
  \item[Direct condition] A writer uses an LLM chat interface with the same
  underlying model and a matched generation budget.
  \item[Plan condition] The writer uses the five-round dialogue and L1--L3
  blueprint, but axiom retrieval, identifiers, relations, and mapped-work
  context are withheld.
  \item[Axiom condition] The writer uses the complete \system{} workflow.
\end{description}

The comparison addresses four questions: (RQ1) Does the dialogue-and-planning workflow
improve independent ratings of long-range coherence relative to direct
generation? (RQ2) Does axiom grounding add thematic integration or useful idea
exploration beyond planning alone? (RQ3) How do the conditions affect perceived
agency, ownership, workload, and satisfaction? (RQ4) Which visible artifacts
do writers accept, edit, replace, or ignore?

\subsection{Study design}

A preregistered within-participant counterbalanced study would ask writers to
develop comparable short-fiction briefs under all three conditions. Assignment
of matched briefs and condition order should be randomized and counterbalanced,
using different briefs for each writer to limit carryover. Model version,
temperature, context window, time allowance, and total generation budget should
be held constant. Recruitment should include writers
with different levels of experience and should record genre familiarity and
prior use of generative tools. Sample size should be fixed before recruitment
using a power analysis tied to the primary contrast and analysis model.

The primary artifact should include both a completed outline and a substantial
prose excerpt so that planning quality and textual realization can be rated
separately. Direct-condition writers would produce the same deliverables using
their own workflow, without an imposed L1--L3 hierarchy. Writers should be
allowed to edit all outputs; otherwise the study
would evaluate model generation rather than human--AI collaboration.

\subsection{Measures and analysis}

Independent literary judges, blinded to condition and authorship, should rate
structural coherence, thematic integration, character development,
distinctiveness, and interpretive richness using the anchored rubric in
Table~\ref{tab:rubric}. Multiple judges per artifact are required, with
inter-rater reliability reported before aggregation. Writer-facing measures
should include the Creativity Support Index \citep{cherry2014csi}, perceived
ownership and agency items, workload, and structured interviews.

Interaction logs should record accepted and rejected axiom candidates, edits to
each blueprint layer, regeneration requests, and time per stage. These process
measures can reveal whether the scaffold is actively used rather than merely
present. The primary analysis should use mixed-effects models with participant
and prompt as random effects, report effect sizes and uncertainty intervals,
and correct planned multiple comparisons. Qualitative data should be coded by
at least two researchers with disagreements resolved and the coding procedure
reported.

The study should preregister exclusion rules and primary outcomes, archive the
prompts and model settings, and retain enough interaction logs to reproduce
condition exposure. Literary judges should assess outputs without seeing the
system's own heuristic scores.

\subsection{Proposed rubric and ablation record}
\label{sec:rubric}

The following instruments are proposed for a future study and have not been
validated. Artifact criteria use a five-point scale with passage or outline
evidence. Judges receive the same kind of brief and anonymized output in every
condition, together with a fixed comparator set for distinctiveness judgments;
they infer thematic commitments from these materials, without
seeing selected axioms. Writer agency is measured separately through self-report
and interviews. Axiom realization is a condition-specific process measure.

\begin{table}[t]
\centering
\small
\sloppy
\begin{tabularx}{\linewidth}{@{}>{\raggedright\arraybackslash}p{0.20\linewidth}Y Y@{}}
\toprule
\textbf{Dimension} & \textbf{Low anchor (1)} & \textbf{High anchor (5)} \\
\midrule
Structural coherence & Major causal, temporal, or character contradictions;
the outline cannot explain key transitions & Causal and temporal transitions are
legible, with purposeful exceptions that the artifact supports \\
Thematic integration & Themes suggested by the brief and text are stated without
consequences in events and choices & Thematic commitments recur through
concrete choices while remaining open to interpretation \\
Character development & Character changes are asserted or unrelated to the
central conflict & Changes are motivated, legible across beats, and carry a cost \\
Distinctiveness & The premise and realization rely on generic or familiar
patterns without a marked angle & The work offers a specific angle that judges
can distinguish from the provided comparators \\
Interpretive richness & The artifact closes interpretation with explanation or
leaves symbols unmotivated & Multiple readings are supported by textual or
structural evidence without requiring one official interpretation \\
Writer agency & The writer reports accepting suggestions because alternatives
were unavailable or opaque & The writer can explain what was accepted, changed,
or rejected and why \\
\bottomrule
\end{tabularx}
\caption{Proposed human-rating anchors. These anchors are a study instrument,
not validated measures or system scores.}
\label{tab:rubric}
\end{table}

An ablation record should preserve matched writing briefs and common prompt
components, model identifier, temperature, target length, and user time allowance.
Total token and model-call budgets should be matched across conditions, including
planning and retrieval-hypothesis generation, with actual usage archived.
Complete condition-specific prompts must be retained: identical prompts would
remove the intended manipulation, and budget matching does not equate semantic
content. The Axiom condition records the selected axiom IDs, roles, relation
edges shown, and edits to each creative core. The Plan condition records the
same blueprint artifacts with the knowledge fields withheld. The Direct
condition records the initial brief and the
same final prose budget without exposing the scaffold. All three conditions
should archive the writer's edits and the final artifact; otherwise the study
could confound the support provided with unrecorded differences in human labor.

The minimum analysis unit is a writer--prompt pair. Planned contrasts should
be specified before looking at outcomes: Plan versus Direct estimates the
combined dialogue, interface, and planning contribution; it does not isolate
hierarchy alone. Axiom versus Plan estimates the incremental
grounding contribution. Secondary analyses can examine whether writers with
different experience levels use the visible axiom fields differently. A
successful manipulation check should verify that participants noticed and
understood the selected propositions; a null manipulation check would make an
effect difficult to interpret.

\section{Discussion}
\label{sec:discussion}

\subsection{A shared vocabulary, not a quality oracle}

The principal design value of literary axioms is that they can be pointed to,
disputed, combined, and replaced. This supports a form of coordination that a
latent model representation cannot offer directly. A writer can say that a
suggested conflict realizes the wrong proposition, or that two selected
propositions should remain unresolved. Such exchanges may support reflection
even when the system's recommendation is rejected.

The same explicitness can create anchoring. A retrieved proposition may narrow
exploration, canonical work mappings may confer undeserved authority, and
numeric heuristics may appear objective. Interface design must therefore show
provenance and context, permit browsing beyond top-ranked results, and avoid
presenting scores as certification. The future evaluation should test these
possible harms as well as intended benefits.

\subsection{Scope of the current prototype}

The prototype establishes that the proposed workflow can be represented and
executed in software. It does not establish causal improvement in writing,
learning, or creative agency. Current generation behavior also depends on the
configured third-party model, and prompt adherence can vary across providers.
There is no automatic provider fallback, general blueprint branching, learned
writer-preference model, or validated confidence calibration in the reported
implementation.

The knowledge base is limited by its source selection, extraction process,
translations, and taxonomy. Coverage of 149 mapped works cannot represent the
diversity of global literary traditions. Context types and domain labels make
some scope assumptions explicit, but do not establish reduced cultural bias.
The framework should grow through documented
provenance, multilingual review, and disagreement-preserving annotation rather
than by treating curation as consensus.

\subsection{Ethics, authorship, and data governance}

Writers should know which model provider receives their text and whether that
provider retains it. A production deployment needs explicit retention controls,
export and deletion mechanisms, and a policy for sensitive drafts. The
real-event module requires stronger safeguards than a generated
identifiability score; users need warnings about consent, living persons,
confidential information, and legal review.

Generated prose may reproduce training-data patterns or stereotypes. Axiom
grounding does not eliminate those risks, and mappings to canonical works can
amplify a narrow canon. Human authors remain responsible for review,
attribution, and compliance with venue or publisher disclosure rules. Future
studies involving writers must receive appropriate ethics review and informed
consent before data collection.

\section{Conclusion}

\system{} explores a specific alternative to direct prompt-to-text writing: an
explicit, editable knowledge layer that connects a writer's intent to narrative
planning and subsequent prose generation. The implemented prototype combines a
five-round dialogue, semantic and relation-aware axiom matching, a three-level
blueprint, scene-level writing operations, real-event abstraction, and advisory
diagnosis. Its contribution is the architecture and the inspectable path it
creates from selected propositions to creative artifacts. Determining whether
that path improves coherence, exploration, or writer agency requires the
controlled comparison outlined in this paper.

\section*{Artifact Availability}

An online demonstration of the current hosted system is available at
\url{https://www.incipit.vip/en-US}. It allows readers to inspect and try the
reported interaction workflow, subject to service availability, account
requirements, provider configuration, and changes to the hosted deployment.
The live service is not a versioned reproducibility artifact: the source code,
curated data, prompt templates, and deployment configuration remain private at
the time of writing, and no archival identifier or license is claimed for them.
A future release should include versioned software and data, prompt templates,
schema migrations, evaluation materials, provenance documentation, and
licenses after source and rights review.

\section*{Author Contributions}

Qiang Liu: conceptualization, methodology, system design and implementation,
and writing--original draft. Chunyi Zhao: domain-informed critical review from
a literary-studies perspective and writing--review and editing.

\appendix

\section{Implementation Manifest}
\label{app:manifest}

This appendix records implementation facts that are easy to lose when the
prototype evolves. It is a compact manifest for the version described in this
paper, not a claim that the private repository is a permanently reproducible
release.

\begin{table}[!htbp]
\centering
\small
\begin{tabularx}{\linewidth}{@{}p{0.28\linewidth}Y@{}}
\toprule
\textbf{Item} & \textbf{Verified value or boundary} \\
\midrule
Knowledge records & 1,455 axiom records; 472 typed relationships; 149 mapped
works; values audited in the companion paper \\
Direction dialogue & Five ordered rounds: story core, core conflict,
differentiation, expression style, confirmation \\
Semantic retrieval & Cosine search; top-$k=5$ per hypothesis; minimum score
0.2; candidate union capped at 50 \\
Assembly context & First 30 candidates plus relations whose endpoints are in the
candidate pool \\
Weak-retrieval branch & Up to three proposed records, status \texttt{proposed},
retrieval retry if records were added or retrieval failed; seed-only evaluation if retrieval remains unavailable \\
Blueprint lengths & Defaults use Chinese-character-oriented length units,
not a verified English-word conversion. Short: 8,000--15,000 and 3--5 chapters; medium:
20,000--60,000 and 8--15; long: 80,000--200,000 and 20--40 \\
Blueprint schema & JSONB persistence with Pydantic validation; five structure
labels; chapter reference membership checked for a nonempty L2 beat set \\
Budget normalization & Chapter and scene target, minimum, and maximum counts are
recomputed from the selected length profile; bounds default to 80\% and 120\% \\
Diagnosis input & Project content is truncated to the first 6,000 characters;
classic-work context is empty; normal selected-record snapshots omit context type \\
LLM recovery & One JSON repair request after a parse failure; no automatic
provider failover \\
Embedding constraint & ORM vector column is 1,024-dimensional; provider and
database dimensions must be coordinated \\
\bottomrule
\end{tabularx}
\caption{Implementation manifest derived from the checked-in backend. Values are
configuration and boundary facts, not performance measurements.}
\label{tab:manifest}
\end{table}

\clearpage
\section{Selected Prompt and Schema Excerpts}
\label{app:prompts}

The complete templates remain in the private source tree. The excerpts below
show the kinds of constraints supplied to the model and the machine-readable
keys that are subsequently checked. They are included to make the design more
specific without implying that a prompt alone controls model behavior.

\begin{table}[!htbp]
\centering
\small
\begin{tabularx}{\linewidth}{@{}p{0.18\linewidth}Y Y@{}}
\toprule
\textbf{Template} & \textbf{Selected instruction} & \textbf{Checked result} \\
\midrule
S1 understand & Express the core conflict as two opposing forces; infer missing
fields cautiously & \texttt{core\_conflict}, \texttt{characters},
\texttt{setting}, \texttt{emotional\_tone}, \texttt{thematic\_hints} \\
S2 hypothesize & Return 5--10 differentiated propositional hypotheses, each
15--40 words & At least three hypotheses before continuing \\
S4 assemble & Select primary, supporting, and form roles; use tensions for
narrative momentum; avoid direct contradiction relationships & Selected axiom UUID membership
is checked against retrieved candidates \\
L1 setting & Match core-title and manifestation arrays to input cores; derive
characters, genre, narrative, tone, and themes & L1 model fields and normalized
creative-core records \\
L2 structure & Produce 5--8 beats with IDs \texttt{beat-1}, \texttt{beat-2},
... and a complete character arc & Enumerated structure type and emotional levels
in $[0,1]$ \\
L3 outline & Associate every chapter with at least one beat and assign scene
word budgets & Beat membership when L2 is nonempty; empty chapter references
permitted; budget normalization \\
Diagnosis & Return five dimensions, findings, comparable works, and actionable
suggestions; scores range 0--100 & JSON is stored as a diagnosis history entry;
no expert calibration \\
\bottomrule
\end{tabularx}
\caption{Prompt constraints and the corresponding implementation checks.}
\label{tab:prompt-excerpts}
\end{table}

The templates also contain a language policy that keeps JSON keys, enum values,
IDs, and other machine-readable fields in English while allowing content to be
generated in the selected language. This policy is operationally important for
validation: translating an identifier or enum value would make a literary
choice look like a schema error.

\clearpage
\section{Proposed Study Data Dictionary}
\label{app:data-dictionary}

The future evaluation should archive enough information to distinguish the
writer's choices from system defaults. The following fields are proposed; none
is currently available as a study dataset.

\begin{table}[!htbp]
\centering
\small
\begin{tabularx}{\linewidth}{@{}p{0.25\linewidth}Y@{}}
\toprule
\textbf{Record} & \textbf{Fields to archive} \\
\midrule
Participant & Pseudonymous ID, writing experience, genre familiarity, prior
generative-tool use, consent and withdrawal status \\
Prompt instance & Brief text, topic, condition, counterbalance position, target
length, time allowance, and model/provider manifest \\
Axiom exposure & Candidate IDs shown, selected IDs, roles, relations displayed,
proposed/reviewed record status, and writer edits \\
Blueprint history & L1/L2/L3 snapshots, patch operations, chapter regeneration
requests, and timestamps relative to the session start \\
Prose artifact & Final outline, generated draft, writer edits, final excerpt,
word counts, and whether the writer accepted or rejected the artifact \\
Judge record & Blind judge ID, rubric scores, cited evidence, adjudication
status, and any missing-data code \\
Process record & Turn count, option selection, custom responses, diagnosis
requests, and explicit user corrections \\
\bottomrule
\end{tabularx}
\caption{Proposed archival fields for a future preregistered comparison. Data
collection requires ethics review, consent, and a retention policy.}
\label{tab:data-dictionary}
\end{table}

\clearpage
\bibliographystyle{plainnat}

\end{document}